\documentclass[sigconf]{acmart}
\usepackage{booktabs} 

\usepackage{booktabs}
\usepackage{enumitem}

\newcommand{\ignore}[1]{}

\setcopyright{none}
\renewcommand\footnotetextcopyrightpermission[1]{} 
\begin{document}

\title{The Simplest Thing That Can Possibly Work:\\ Pseudo-Relevance Feedback Using Text Classification}

\author{Jimmy Lin}
\affiliation{\vspace{0.1cm}
  \department{David R. Cheriton School of Computer Science}
  \institution{University of Waterloo}
}
\email{jimmylin@uwaterloo.ca}

\begin{abstract}
Motivated by recent commentary that has questioned today's pursuit of ever-more complex models and mathematical formalisms in applied machine learning and whether meaningful empirical progress is actually being made,
this paper tries to tackle the decades-old problem of pseudo-relevance feedback with ``the simplest thing that can possibly work''.
I present a technique based on training a document relevance classifier for each information need using pseudo-labels from an initial ranked list and then applying the classifier to rerank the retrieved documents.
Experiments demonstrate significant improvements across a number of newswire collections, with initial rankings supplied by ``bag of words'' BM25 as well as from a well-tuned query expansion model.
While this simple technique draws elements from several well-known threads in the literature, to my knowledge this exact combination has not previously been proposed and evaluated.
\end{abstract}

\maketitle

\section{Introduction}

The breakneck pace of advances in machine learning, particularly deep learning applied to vision, speech, and text processing tasks, has recently prompted a number of researchers to urge caution and the need for self reflection.
Sculley et al.~\cite{Sculley_etal_2018} and Lipton and Steinhardt~\cite{Lipton:1807.03341v2:2018} represent two recent commentary along these lines.
The trend of increasingly complex models with poor ablation studies to attribute gains, coupled with the use of mathematics to obfuscate or to impress reviewers, has put empirical research on shaky footing.
In a similar vein, I've recently expressed skepticism about whether neural ranking models actually improve over existing models, at least absent large amounts of behavioral training data~\cite{Lin_SIGIRForum2018}.

This paper tackles the decades-old problem of pseudo-relevance feedback.
I am guided by the advice of Ward Cunningham (inventor of the wiki), which is to ask yourself, ``What's the simplest thing that could possibly work?''
Here's the answer I came up with:

Given a standard {\it ad hoc} retrieval setup, a ranking model produces a ranked list $H$ with respect to an information need represented by query $Q$; this is referred to as the base run.
Following the general setup of pseudo-relevance feedback, let's assume that the first $r$ hits are relevant, i.e., \texttt{H[:r]} in Python's array slice notation.
Let's further assume that the last $n$ hits of the ranked list are not relevant, i.e., \texttt{H[-n:]} in Python's array slice notation.
These $r+n$ documents with their pseudo labels are then used to train a text classifier over the \textit{tf-idf} representations of the document terms.
This paper explores logistic regression, SVMs, and a simple ensemble.
The trained classifier is then applied to score all documents in the base run:\ a new document score comprised of a linear interpolation between the initial retrieval score and the classifier score is then used to create the final ranked list (for evaluation).

Experiments on four newswire collections show that this simple technique yields significant increases in effectiveness over a base run from ``bag of words'' BM25 as well as a base run that already exploits pseudo-relevance feedback via RM3.
The latter result suggests that the proposed technique provides additive improvements on top of a strong baseline.

\section{Prior Work}

The obvious retort to this ``simplest thing that can possibly work'' is that, even if it works, it can't possibly be novel!
While the literature does contain reports of similar ideas, this exact combination to my knowledge has not been tried before.

The idea of treating document ranking as a binary classification problem, to distinguish relevant from non-relevant documents, has a long history, dating back to the binary independence retrieval (BIR) model of Robertson and Spark Jones~\cite{Robertson_SparkJones_1976}.
In the modern parlance of learning to rank, this is commonly known as a {\it pointwise model}~\cite{LiuTY_FnTIR2009}.
As an early example, Nallapati~\cite{Nallapati04} used logistic regression and SVMs for document ranking with features based on {\it tf}, {\it idf}, and statistics derived from their combination.

Relevance feedback in IR systems dates back to the 1960s~\cite{Rocchio_1971} and the idea of using pseudo labels for relevance feedback dates back to at least the late 1970s~\cite{Croft_Harper_1979}; a nice historical overview is offered by Ruthven and Lalmas~\cite{Ruthven_Lalmas_2003}.
According to Buckley~\cite{Buckley_TRECbook}, early attempts at pseudo-relevance feedback were unsuccessful because of small collection sizes, and it wasn't until around TREC-2 that researchers demonstrated positive results.
Today, the effectiveness of pseudo-relevance feedback is well established in the literature.

In most setups, including the popular RM3 approach widely used today~\cite{Abdul-Jaleel04}, the first $k$ hits of an initial ranked list are assumed to be relevant.
Some analysis (varies by approach) is applied to these pseudo-relevant documents to generate an expanded query that is then used to produce a final ranked list.
Typically, this class of methods only exploits pseudo-positive labels.
Far less popular, the use of pseudo-negative labels dates back to at least 2003~\cite{YanRong_etal_2003} (the earliest reference that I could find), albeit in the context of multimedia retrieval.
A more recent example is the use of pseudo-negative labels by Raman et al.~\cite{RamanKarthik_etal_ECIR2010} to extract better query expansion terms.
Cormack et al.~\cite{Cormack_etal_IRJ2011} used pseudo-negative labels to train spam classifiers in a distantly-supervised manner.
Note that while there is literature on how to select good expansion terms using \textit{supervised} machine learning techniques~\cite{Cao:2008:SGE:1390334.1390377}, my approach is completely {\it unsupervised}.

There is a thread of research putting together relevance feedback with text classification:\ Cormack and Mojdeh~\cite{Comarck_Mojdeh_TREC2009} applied logistic regression classifiers for relevance feedback in TREC 2009.
Grossman and Cormack~\cite{GC} later proposed a variant where classifiers were trained on relevance judgments on a different collection for the same information need; this work was successfully reproduced by Yu et al.~\cite{Yu_etal_ECIR2019}.
Xu and Akella~\cite{Xu_Akella_SIGIR2008} described an active relevance feedback approach using logistic regression.
All these cases, however, took advantage of human relevance judgments and did not consider pseudo judgments.
I argue that this distinction, while perhaps obvious in retrospective, is quite important---historically, the idea of using pseudo labels came at least a decade after the introduction of relevance feedback, and its empirical value wasn't demonstrated until many years after that (see above).
Yan et al.~\cite{YanRong_etal_2003} described a technique quite similar to what I propose here, expect applied to multimedia retrieval.

Pulling all the pertinent characteristics together---use of both positive and negative pseudo-labels to train text classifiers in a pseudo-relevance feedback setup for {\it ad hoc} retrieval---I assert that this paper is the first to propose such a technique and present experimental results on a number of modern test collections.

\section{Experimental Setup}
\label{impl}

All experiments were conducted using Anserini~\cite{Yang_etal_JDIQ2018}, an open-source IR toolkit built on top of Lucene.\footnote{Commit id \texttt{9548cd6}, dated 01/19/2019.}
Anserini provides convenient tools to dump out raw \textit{tf-idf} document vectors for arbitrary documents, which was used to extract feature vectors for the top $r$ and bottom $n$ documents from the base run for each topic.
As part of preprocessing, all feature vectors were converted to unit vectors by $L_2$ normalization.

For each topic, a training set was created from the documents corresponding to the positive and negative pseudo labels, as described in the introduction.
Feature vectors were then fed to the Python package \texttt{scikit-learn}~\cite{pedregosa2011scikit} (v0.20.1).
I tried three different models:\ logistic regression (LR), SVMs with a linear kernel, and an ensemble of the two using simple score averaging.
In each case, the trained classifier for each topic was then applied to all documents in the ranked list for that topic from the base run.
Documents in the base run were reranked using a linear interpolation between retrieval and classifier scores.

Experiments were conducted on four newswire collections:

\begin{itemize}[leftmargin=*]

\item Robust04:\ TREC Disks 4 \& 5 (excluding Congressional Record) with topics and judgments from the TREC 2004 Robust Track.

\item Robust05:\ The AQUAINT Corpus, with topics and judgments from the TREC 2005 Robust Track.

\item Core17:\ The New York Times Corpus, with topics and judgments from the TREC 2017 Common Core Track.

\item Core18:\ The Washington Post Corpus, with topics and judgments from the TREC 2018 Common Core Track.

\end{itemize}

\noindent The primary test collection was Robust04, to take advantage of baselines and comparisons I've previously established~\cite{Lin_SIGIRForum2018}.
The other newswire collections provide some indication of the generality of the technique, at least for documents of the same genre.
Note that web collections were not considered because most were created with shallow pools, which make them inappropriate for studies on query expansion due to prodigious amounts of missing judgments; see Yang and Lin~\cite{Yang_Lin_ECIR2019} for an example analysis.

My proposed technique has three parameters:\ $r$, the number of pseudo-positive labels, $n$, the number of pseudo-negative labels, and $\alpha$, the interpolation weight.
Preliminary exploration showed $n=100$ to be a good setting (relatively insensitive) and that $r \in \{10, 20, 30\}$ seemed to be good choices.
For the interpolation parameter, all values between $0.0$ and $1.0$ in tenth increments were tried.
All parameter tuning was accomplish via cross validation.

\begin{table}[t]
\centering
\begin{tabular}{rlllll}
\toprule
 & Condition & AP & {\it p}-value\\ 
\toprule
   & {\it Baseline BM25 (5-fold)} \\
1  & BM25                        & 0.2531 &  \\
2  & BM25 + LR                   & 0.2734 &  2.67 $\times 10^{-7}$ \\
3  & BM25 + SVM                  & 0.2685 &  1.29 $\times 10^{-8}$ \\
4  & BM25 + ensemble             & 0.2724 &  2.71 $\times 10^{-8}$ \\
\midrule
   & {\it Default RM3 parameters (5-fold)} \\
5  & BM25 + RM3         & 0.2903 & \\
6  & BM25 + RM3 + LR   & 0.3002 & 0.0001225 \\
7  & BM25 + RM3 + SVM  & 0.2986 & 2.56 $\times 10^{-5}$ \\
8  & BM25 + RM3 + ensemble  & 0.2998 & 1.34 $\times 10^{-5}$ \\
\midrule
   &  \multicolumn{3}{l}{{\it RM3 cross-validation: 2-fold from Paper 1}} \\
9  & BM25 + RM3         & 0.2987 & \\
10  & BM25 + RM3 + LR   & 0.3035 & 0.0241963 \\
11  & BM25 + RM3 + SVM  & 0.3031 & 0.0015774 \\
12  & BM25 + RM3 + ensemble & 0.3023 & 0.0112373 \\
\midrule
    &  \multicolumn{3}{l}{{\it RM3 cross-validation: 5-fold from Paper 2}} \\
13  & BM25 + RM3         & 0.3033 & \\
14  & BM25 + RM3 + LR   & 0.3092 & 0.0105262 \\
15  & BM25 + RM3 + SVM  & 0.3096 & 1.02 $\times 10^{-5}$ \\
16  & BM25 + RM3 + ensemble  & 0.3082 & 0.0018256 \\
\midrule
17 & Paper 1                           & 0.2720 \\
18 & Paper 2                           & 0.2971 \\
19 & NPRF                              & 0.2904 \\
20 & Best TREC (\texttt{pircRB04t3})   & 0.3331 \\
\bottomrule
\end{tabular}
\vspace{0.2cm}
\caption{Effectiveness of pseudo-relevance feedback using text classification on Robust04.\label{results:robust04}}
\label{table:compare}
\vspace{-0.7cm}
\end{table}

\begin{table*}[t]
\centering
\begin{tabular}{rlllllll}
\toprule
 &  & \multicolumn{2}{l}{{\bf Robust05}} & \multicolumn{2}{l}{{\bf Core17}} & \multicolumn{2}{l}{{\bf Core18}}\\
 & Condition & AP & {\it p}-value & AP & {\it p}-value & AP & {\it p}-value \\ 
\toprule
1 & BM25                  & 0.2031 &          & 0.1977 &          & 0.2491 \\
2 & BM25 + LR             & 0.2457 & 0.000203 & 0.2318 & 0.002472 & 0.2791 & 0.026064 \\ 
3 & BM25 + SVM            & 0.2404 & 0.000720 & 0.2228 & 0.004935 & 0.2798 & 0.000309 \\ 
4 & BM25 + ensemble       & 0.2446 & 0.000395 & 0.2298 & 0.002677 & 0.2743 & 0.034362 \\ 
\midrule
5 & BM25 + RM3            & 0.2602 &          & 0.2682 &          & 0.3147  \\
6 & BM25 + RM3 + LR       & 0.2820 & 0.001307 & 0.2882 & 0.001146 & 0.3214 & 0.313905 \\
7 & BM25 + RM3 + SVM      & 0.2798 & 0.001533 & 0.2855 & 0.000178 & 0.3273 & 0.011007 \\
8 & BM25 + RM3 + ensemble & 0.2814 & 0.001549 & 0.2880 & 0.000318 & 0.3286 & 0.030315 \\
\midrule
9 & TREC best (automatic) & 0.3096 &          & 0.2752 &          & 0.2761 & \\
\bottomrule
\end{tabular}
\vspace{0.2cm}
\caption{Effectiveness of pseudo-relevance feedback using text classification on Robust05, Core17, and Core18.\label{results:other}}
\vspace{-0.5cm}
\end{table*}

\section{Results}

Experimental results on Robust04 in terms of average precision at rank 1000 are shown in Table~\ref{results:robust04}.
In the rows, ``LR'', ``SVM'', and ``ensemble'' refer to different text classification models discussed in Section~\ref{impl}.
Rows 2--4 report my technique applied to a ``bag of words'' BM25 run ($k_1=0.9$, $b=0.4$).
Rows 6--8 report my technique applied to RM3 (using default parameters from the open-source Indri Search Engine) on top of a BM25 base run.
In both cases results are reported for parameter tuning ($r$, $n$, and $\alpha$) using five-fold cross validation.
For these and all subsequent experiments, statistical significance of metric differences was assessed using a paired two-tailed {\it t}-test.
Cognizant of the dangers of multiple-hypothesis testing, the right column reports the exact {\it p}-values, which allows readers to make corrections for multiple hypothesis testing as they feel appropriate.
Results show that my proposed technique (all models) unequivocally improves average precision (even, for example, after applying a Bonferroni correction).

It is worth pointing out that this technique only acts as a reranker---it cannot ``produce'' any relevant document that was not in the base run.
Thus, the improvements over BM25 come solely from bringing relevant documents into higher ranks.
Since RM3 performs query expansion in a second round retrieval, it is able to retrieve more relevant documents; my technique can further improve the ranks of those documents.

In the above conditions no effort was made to tune parameters for the base runs.
The next set of experiments examined whether pseudo-relevance feedback using text classification can further improve over well-tuned base runs, building on the comparisons to ``Paper 1'' and ``Paper 2'' in my recent article~\cite{Lin_SIGIRForum2018}.
In Table~\ref{results:robust04}, rows 9 and 13 replicate those tuned baselines,\footnote{Note that these results are slightly higher than those reported in the SIGIR Forum article due to improvements made after publication.} on top of which another round of reranking was applied.
Results are shown in rows 10--12 for Paper 1 and rows 14--16 for Paper 2.
Based on the {\it p}-values, the improvements would be considered significant, although not so after a Bonferroni correction, except for SVM.

It should be emphasized that not only does pseudo-relevance feedback using text classification demonstrate improvements over different base runs, but also that the reported metrics are quite high in absolute terms.
For reference, the best results reported in Paper~1 and Paper 2 are copied in rows 17 and 18 for reference.
Also included are results from NPRF, a recently-proposed neural approach to pseudo-relevance feedback~\cite{Li_EMNLP2018} (row 19).
The base runs already surpass the highest effectiveness reported in all these papers, and my technique further increases retrieval effectiveness.
In other words, pseudo-relevance feedback using text classification is both better and simpler.
Nevertheless, its effectiveness still falls short of \texttt{pircRB04t3}, the best run from TREC 2004 (row 20).

Results from the three other collections are shown in Table~\ref{results:other}, organized in the same manner as Table~\ref{results:robust04}.
Reported are the applications of my technique on top of ``bag of words'' BM25 as well as BM25 with RM3 (both with default parameters, comparable to rows 1 and 5 in Table~\ref{results:robust04}).
As with the previous experiments, the parameters $r$, $n$, and $\alpha$ were tuned via five-fold cross-validation.
I have not extended the Robust04 tuning experiments to these collections, so additional points of comparison are not available.
For reference, row 9 presents the best automatic run at that particular year's TREC.

These results are consistent with the results on Robust04, providing evidence that my technique generalizes across different collections.
For Robust05 and Core17, all gains appear to be statistically significant, although for Core18 only the SVM models are (despite the fact that metric increases are comparable in magnitude for logistic regression).
In absolute terms, for Core17 and Core18, all results on BM25 + RM3 are higher than the best automatic TREC run submitted that year.

Across all four collections, logistic regression in most cases is slightly better than SVMs in terms of effectiveness, but the gains from logistic regression are not significant for all collections.
Results further suggest that an ensemble based on simple score averaging yields no benefit over individual models.

\section{Discussion}

Given the simplicity and effectiveness of my proposed technique for pseudo-relevance feedback using text classification, the obvious question is:\ Where are the gains coming from?
While it may be no surprise that the technique improves a ``bag of words'' BM25 base run, it also improves RM3, a base run that already exploits pseudo-relevance feedback.
The gains, albeit smaller, are significant (across all collections, at least for SVMs).
This suggests that my technique is extracting additional relevance signal beyond what RM3 can identify.

The explanation, I believe, lies in what Diaz~\cite{Diaz_CIKM2015} calls score regularization in ranked retrieval, which is the idea that closely-related documents should have similar scores.
This itself is a restatement of the decades-old cluster hypothesis~\cite{Jardine_vanRijsbergen_1971}, which is the observation that relevant documents tend to share similar content (i.e., cluster in document space). 
Thus, effectiveness gains come from breaking the independence assumption in ranking, which still holds in RM3.
Another way to summarize these results is that these observed effectiveness improvements are {\it additive} with respect to RM3.
This is an important finding because the question of additivity has significant bearing on the methodology of empirical research in information retrieval.
Previous work~\cite{Armstrong:2009:IDA:1645953.1646031,Kharazmi_etal_TOIS2016} has found that many techniques only improve over weak baselines:\ when applied to stronger baselines, observed gains disappear.
In other words, the relevance signals that these techniques exploit are subsumed by the stronger baselines.
In this context, the regularization effect potentially explains {\it why} improvements are additive:\ it taps a different source of signal.

\begin{figure}[t]
\includegraphics[width=0.85\linewidth]{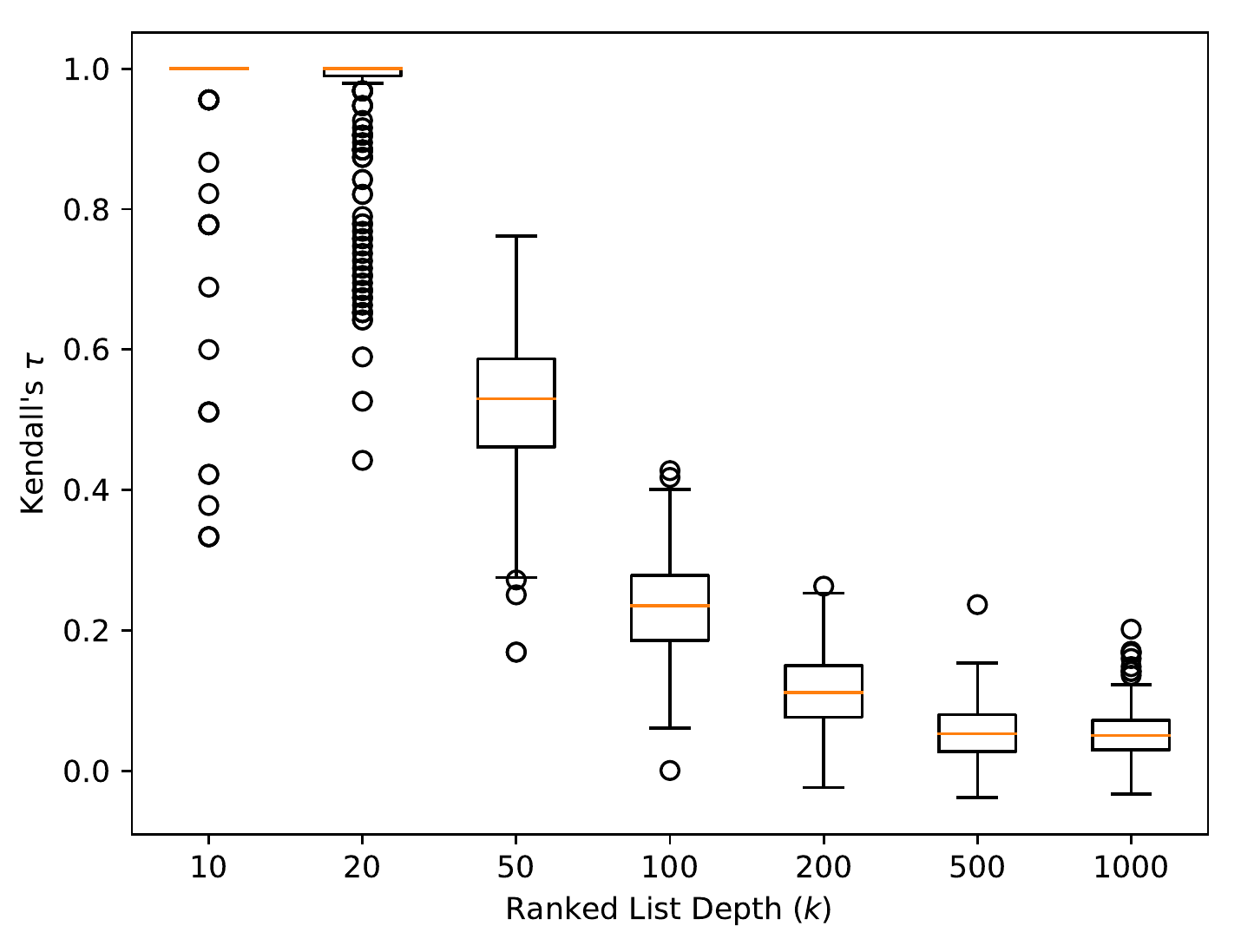}
\vspace{-0.2cm}
\caption{Results of a rank correlation analysis on Robust04, comparing BM25 vs.\ BM25+RM3 base runs at different $k$.\label{fig:tau-analysis}}
\vspace{-0.3cm}
\end{figure}

A closer look at the base and reranked runs reveals interesting observations.
For example, compare row 13 vs.\ row 15 in Table~\ref{results:robust04} as a representative case.
On a per-topic basis, SVMs helped 70 topics, arbitrarily defined as AP increases $>0.01$;
SVMs hurt 36 topics, arbitrarily defined as AP decreases $>0.01$;
for the remaining topics (143) SVMs didn't make much of a difference (AP either remained unchanged or changed little).
On the whole, it appears that my technique yields an overall improvement by helping only a relatively small fraction of topics, which is why we observe statistical significance even though the magnitude of the improvements is small.
Nevertheless, the technique does appear to be reshuffling the ranked lists quite dramatically:\
Figure~\ref{fig:tau-analysis} compares rank correlation (measured using Kendall's $\tau$) between the base and final runs at ranks $\{10, 20, 50, 100, 200, 500, 1000\}$.
The box-and-whiskers plots show the per-topic distribution of the Kendall's $\tau$ values.
As expected, the early ranks change little since those documents are used as pseudo-positive labels.
However, beyond the early ranks, the ``before'' and ``after'' results are quite different.
Interestingly, in most cases, this large reshuffling does not appear to change AP much.

\section{Conclusions}

One common thread in recent commentary referenced in the introduce is that complexity (in terms of models, parameter estimation, etc.)~potentially obscures understanding, especially in the absence of rigorous ablation studies.
Simplicity is a virtue, and simple yet effective techniques should be the most preferred type of solution overall.
I believe my technique fits this description.

\section*{Acknowledgments}

This research was supported by the Natural Sciences and Engineering Research Council (NSERC) of Canada.

\bibliographystyle{ACM-Reference-Format}
\bibliography{lr-arxiv}

\end{document}